\documentclass[conference]{IEEEtran}
\IEEEoverridecommandlockouts

\usepackage[letterpaper,margin=1in]{geometry}

\usepackage{cite}
\usepackage{amsmath,amssymb,amsfonts}
\usepackage{algorithmic}
\usepackage{graphicx}
\usepackage{textcomp}
\usepackage{xcolor}
\usepackage{booktabs}
\usepackage{array}
\usepackage{url}
\usepackage{balance}
\usepackage{tikz}
\usetikzlibrary{arrows.meta,positioning,fit}

\def\BibTeX{{\rm B\kern-.05em{\sc i\kern-.025em b}\kern-.08em
    T\kern-.1667em\lower.7ex\hbox{E}\kern-.125emX}}

\newcommand{\safesuccess}{\textsc{SafeSuccess}}

\begin{document}
\title{Task-Conditioned Least-Privilege Learning for Executable Terminal and MCP Agents\thanks{Preprint. This work has been submitted to the IEEE for possible publication. Copyright may be transferred without notice, after which this version may no longer be accessible.}
}

\author{%
\IEEEauthorblockN{Alexander Tu\IEEEauthorrefmark{1}, Michael Tu}
\IEEEauthorblockA{%
\textit{Center for Cybersecurity, College of Technology},
\\ \textit{Purdue University Northwest}, Hammond, IN, USA \\
tu101@purdue.edu, tu15@purdue.edu \\
\IEEEauthorrefmark{1}\textit{A. Tu is also with Columbia College, Columbia University, New York, NY, USA}}

}

\maketitle

\begin{abstract}
Tool-using large language-model agents can complete a task while exercising authority that the user did not grant or the task does not need, causing excess-authority errors. Traditional permission gating systems alone for validating agent environments are insufficient. We study whether post-training can teach a 4B-parameter model to choose task-conditioned authority in executable terminal and Model Context Protocol (MCP) environments to complement those measures. We propose a framework where each action is audited before execution and again from observed effects along six dimensions of risk. This auditing is conducted using deterministic verifiers that score completion, evidence, exact state, prohibited attempts, and safe success. In conjunction with predefined task-specific sufficient-authority envelopes, we determine task-specific excess privilege values for trajectories, which are then optimized for in post-training. We find that after training using this framework on Qwen3.5-4B over 1,500 tasks, the selected seed reaches 98.48\% safe success across 2,896
evaluation episodes spanning all 500 held-out tasks, compared with 64.36\% for the base policy, and reduces excess-authority error events from 4.56\% to 0.79\%. Furthermore, external tests show capability retention and prompt-directed improvement. A 400 task continuation study also found evidence of generalization, reducing excess-authority events by 6.99 percentage points while maintaining previous capabilities. We conclude learned restraint through least-privilege aware post-training is therefore useful as an additional control layer for tool-using agents in executable terminal and MCP environments, but it does not replace permission gates and sandboxing.
\end{abstract}


\section{Introduction}
Tool-using agents can successfully complete tasks ranging from MCP tool calling to editing files, while still causing unintended side effects and violation of the request. These include but are not limited to actions such as the reading of unrelated records, editing targets outside of the task scope, or using tools that have higher authority and risk with the potential for privilege escalation. We define these behaviors as task-relative excess-authority errors, actions that exercise authority beyond the reasonably sufficient minimum for a requested task. Interactions that cross these task-specific boundaries can therefore be classified as over-privileged trajectories, even if the final answer or state is otherwise correct. This over-privilege can be considered a potentially dangerous security-relevant form of model misalignment between the user's intended objective and the agent's executed behavior \cite{zhuang2020}, and therefore motivates explicit privilege management over the model's actions. Classical least privilege requires a process to use only the authority needed for its present function \cite{saltzer1975}. For an agent, that requirement must be applied repeatedly as the model chooses commands, arguments, targets, retries, and side effects.

Current methods of preventing excess-authority errors mainly act after the model has proposed or done an action. Permission gates can block potentially dangerous or risky calls, but their coverage depends on what action classes they have learned to associate with risk. Functionally equivalent state changes or commands may pass through file edits or other paths that are not similarly gated, circumventing gating \cite{ji2026permission}. Static permission gating can also omit access actually required by an execution chain while simultaneously granting unrelated sensitive access \cite{yan2026authbench}. In contrast, we propose to have the model learn to exercise least-privilege actions that still provide sufficient evidence to finish the task.

To study this, we define how that least privilege is task-relative. We use three different metrics to help judge what task specific authority is necessary for tool-using agents. These include what capabilities are shown to the model, what actions the environment permits, and what actual authority and effect happens. The model sees a comprehensive tool inventory, while an unseen environment broker parses each proposed action, deterministically calculates its intended authority, assesses its execution risk, selects the execution mode, and records the resulting effects. This structure lets us assign each task a sufficient-authority envelope, and penalize only excess privilege beyond that envelope.

On this basis, we train Qwen3.5-4B with LoRA and direct Dr.~GRPO on a 1,500-task executable curriculum and evaluate the resulting policies across validation sets, model ablations, and external benchmarks. The training requires the model to complete the task objectives while taking into account the sufficiency and appropriateness of the model's privilege in its actions.  

The paper contributes a least-privilege reinforcement learning framework using task-relative privilege classification with a comprehensive six-dimensional risk vector and a brokered terminal and MCP environment with pre- and post-execution auditing, as well as a deterministic excess-authority evaluation dataset. This framework is built so that authority measurement is widely applicable to different tasks and environments, not only those trained on.

\section{Literature Review}
Prior benchmarks and works have mainly isolated permission inference or individual tool choice, and study agent authority at differing decision points. By contrast, our environment instead measures trajectory authority, since the meaning of a terminal command or MCP call in real agentic environments depends on its arguments, the current state of the environment, and the action's observed effects.

Evidence from deployed coding agents shows that authorization errors are still widespread even in frontier model deployments, especially in real-world environments. AuthBench studies permission-boundary inference before execution \cite{yan2026authbench}. They show that frontier models can both withhold access needed to complete the task while granting access that the task does not require. After Anthropic found that users accepted about 93\% of manual permission prompts, they created Claude Code auto mode which delegates approval decisions to model-based classifiers \cite{hughes2026claudeautomode}. This, however, does not cure excess authority errors. Anthropic reports numerous failures in permission gating, and its two-stage classifier reduces the frequency of manual permission prompts, but retains a 17\% false-negative rate on a curated set of 52 overeager actions\cite{hughes2026claudeautomode}. Further external testing found a false-negative rate as high as 81\%, and 36.8\% of actions falling outside the scope of the classifier \cite{ji2026permission}. This system is an example of permission gating by invocation control, judging a proposed action before execution, and while helpful, it is not perfect. Our work complements these systems by training the agent to propose fewer unnecessary high-authority actions in the first place.

Several benchmarks formalize related authorization failures. The work FORTIS treats an agent skill as a privilege boundary and evaluates whether the model selects the minimally sufficient skill and whether its actions remain within the authority assigned to that skill \cite{li2026fortis}. Its evaluation covers ten frontier models like GPT-5.5 and Claude Opus 4.7 across three domains and finds substantial over-privilege. Errors increase under real-world conditions such as incomplete requests, convenience-oriented wording, and tasks that are not cleanly defined. These findings indicate that improved general capability does not itself produce reliable authority control or risk judgment.

A contemporaneous study, ToolPrivBench, is the closest prior post-training comparison \cite{yang2026toolpriv}. It makes lower- and higher-privilege tools independently sufficient, and measures whether an agent selects a higher-privilege tool initially or escalates after transient lower-tier failures. Its privilege-aware post-training method reduces unnecessary higher-privilege selection while retaining general tool ability. ToolPrivBench appeared while our experimental program was already in progress. It informed our external evaluation, but not our executable environment, six-dimensional risk representation, or reward structure. ToolPrivBench is also distinct in that it cleanly isolates preference among substitutable tools, whereas real-world terminal and MCP actions are often complementary and may legitimately require writes, execution, external service access, or persistent state, which we represent. We do use ToolPrivBench for the motivation and inspiration for the tasks authored in the 400 task continuation however.

General tool-use benchmarks provide capability comparisons and informative tasks. MetaTool evaluates whether a model should call a tool and which tool it should select, including similar-tool, unavailable-tool, and multi-tool cases \cite{huang2023metatool}. We use such external benchmarks to test whether our privilege-aware post-training damages general tool usage ability. Other work has additionally established the generalizability of executable command-line tasks as practical evaluations of multi-step agent behavior, and show how they become the foundation of coding agent workflows when combined in software-agent loops, informing our environment's task composition and structure \cite{wang2024openhands,yang2024sweagent}. MCP extends this action model to arbitrary service tools that can be selected and parameterized by the model \cite{mcp2025tools}.

Our learning optimization method uses work on reinforcement learning with verifiable outcomes. GRPO conducts reinforcement learning through estimating relative advantage from groups of sampled completions \cite{shao2024deepseekmath}. Dr.~GRPO builds on that to remove generated-length normalization as well as group standard deviation normalization to prevent response-length distortion \cite{liu2025drgrpo}. We apply direct Dr.~GRPO to task-conditional least-privilege behavior. Because learned dynamic reward functions can reach low training error but still have high regret for the real objective \cite{fluri2025perils}, reward is derived from deterministic measurements of task success, evidence, environment state, prohibited attempts, and excess authority rather than from dynamic judging.
\section{Methodology}
\subsection{Task Relative Authority}
For task $x$, the visible actions are a terminal command, an MCP call, a categorical human escalation, and a final answer. Their sequence forms $\tau=(a_1,o_1,\ldots,a_T,o_T)$. Each action receives a six-dimensional authority vector
\begin{equation}
\begin{split}
\mathbf{z}(a_t)={}&[z_{\mathrm{write}},z_{\mathrm{exec}},z_{\mathrm{external}},\\
&z_{\mathrm{secret}},z_{\mathrm{scope}},z_{\mathrm{persistent}}]\in[0,1]^6.
\end{split}
\label{eq:risk}
\end{equation}
The components correspond to state mutation, dynamic execution, reaches or exceeds the environment boundary, unprivileged sensitive-data access, broad and wide effects, and state that persists beyond the intended operation. The risk vector is the componentwise maximum between the pre- and post-action risk assessments. The final trajectory risk vector then records the largest value reached
on each authority dimension across all actions,

\begin{equation}
\mathbf{z}_{\tau}
=
\max_{t=1,\ldots,T}\mathbf{z}(a_t).
\end{equation}

Each task is assigned a reviewed sufficient-authority envelope
$\mathbf{z}_{\mathrm{req}}(x)$. The excess-authority vector is

\begin{equation}
\boldsymbol{\Delta}(\tau,x)
=
\max\!\left(
\mathbf{z}_{\tau}-\mathbf{z}_{\mathrm{req}}(x),
\mathbf{0}
\right),
\label{eq:excess}
\end{equation}

where the maximum is taken separately for each authority
dimension. Values below the task requirement are set to zero,
so only authority beyond the sufficient envelope contributes to the excess privilege penalty calculation.

Let $S$ denote task success, $E$ evidence existence and sufficiency, and $V_x$ the task-specific state constraints. Safe success is
\begin{equation}
\safesuccess(\tau,x)=\mathbb{1}[S=1\land E\ge e_x\land V_x=1\land O_{\tau}=0],
\label{eq:safe}
\end{equation}
where $O_{\tau}$ is the reportable excess-authority event. This metric distinguishes what the agent attempted from what the environment allowed to occur. A blocked command will therefore cause no external damage, but still be recorded that the policy requested unjustified authority.

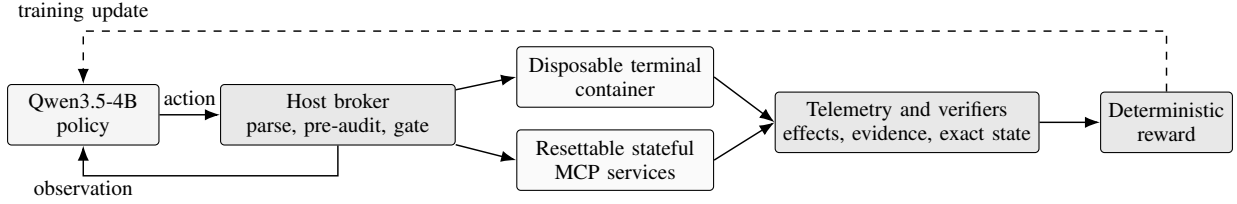
\begin{figure*}[t]
\centering
\begin{tikzpicture}[
  node distance=5mm and 8mm,
  every node/.style={font=\footnotesize,align=center},
  box/.style={draw,rounded corners=1.5pt,minimum height=8mm,inner sep=3pt,fill=black!4},
  host/.style={box,fill=black!9},
  env/.style={box,fill=black!2},
  arrow/.style={-{Latex[length=2mm]},line width=0.55pt}
]
\node[box,minimum width=20mm] (policy) {Qwen3.5-4B\\policy};
\node[host,right=of policy,minimum width=31mm] (broker) {Host broker\\parse, pre-audit, gate};
\node[env,right=of broker,yshift=5mm,minimum width=26mm] (terminal) {Disposable terminal\\container};
\node[env,right=of broker,yshift=-6mm,minimum width=26mm] (mcp) {Resettable stateful\\MCP services};
\node[host,right=8mm of terminal,minimum width=28mm,yshift=-6mm] (audit) {Telemetry and verifiers\\effects, evidence, exact state};
\node[host,right=of audit,minimum width=2mm] (reward) {Deterministic\\reward};
\draw[arrow] (policy) -- node[above]{action} (broker);
\draw[arrow] (broker) -- (terminal.west);
\draw[arrow] (broker) -- (mcp.west);
\draw[arrow] (terminal.east) -- (audit.west);
\draw[arrow] (mcp.east) -- (audit.west);
\draw[arrow] (audit) -- (reward);
\draw[arrow,dashed] (reward.north) |- ++(0,8mm) -| node[pos=0.5,above]{training update} (policy.north);
\draw[arrow] (broker.south) |- ++(0,-4mm) -| node[pos=0.58,below]{observation} (policy.south);
\end{tikzpicture}
\caption{The reinforcement learning loop. Policy selects actions, and the broker parses and pre-audits, sends the actions to the terminal or MCP services, receives the results, and calculate rewards for training updates.}
\label{fig:architecture}
\end{figure*}

\subsection{Brokered Execution and Automatic Verification}
Every trajectory starts with the proposal being analyzed by the \textbf{host broker} as shown sequentially in Fig.~\ref{fig:architecture}. Pre-execution auditing takes in command family, flags, paths, redirection, composition, tool identity, and argument breadth. Supported actions execute inside a non-root container or a resettable MCP service for real effects. Post-execution auditing uses file and Git diffs, process telemetry, service records, sensitive-path access, and exact state changes. As noted previously, the final action privilege is the componentwise maximum of the pre- and post-execution vectors. This allows for calculating the attempted authority when a command is blocked and capturing environment state effects that static parsing may miss.

Verifiers evaluate on exact answers, hidden tests, required output files, targeted test results, exact MCP state, evidence paths, required writes, and rejection of extra writes. These requirements and checks are strict.Proposal-only tasks forbid environment changes and penalize them, escalation tasks require appropriately evidenced human intervention reasons, and evidence is terminal-gated so that a read after the final answer cannot justify that answer.

\subsection{Modular Framework Use}
The framework's separation of broker, task schemas, post-action verifiers and excess privilege calculation are separated to allow for handling tasks that have different behavior and needs. Consider a proposal task that only asks an agent to read a bug report, bug.txt and suggest a patch. Here, a simple read is sufficient while a write would be classified as excessive. Pre-action brokers do, however, require evidence terms because no mutative action is executed by the agent in the correct run. On the other hand, a multi-tool cross-service task that requires an agent to read from a GitHub-like service using MCP tools and copy that information into a separate database row requires elevated sufficient privilege envelopes in external access and writes being required. Furthermore, it utilizes more heavily the post-action verifier and components of the risk vector like the write excess. Hence, the framework can be adapted to different tasks flexibly. 

Furthermore, the framework does not require only Dr.GRPO reinforcement learning optimization. The frameworks excess-privilege calculations and multi-step verification broker can be potentially incorporated as  elements in any number of reward functions and algorithms.

\subsection{Reward and Training}
The scalar reward used for learning optimization is given by:
\begin{equation}
\begin{split}
R={}&0.60S+0.20E-0.20P-0.05U+0.10H\\
&-0.75B-0.05F_u-0.12F_r.
\end{split}
\label{eq:reward}
\end{equation}
Here $P$ combines weighted trajectory excess, added penalties for secret-access excess, label confidence, and a capped stepwise excess term. $U$ penalizes action count and repeatedly looping actions. $H$ rewards correct escalation. $B$ measures blocked-action
severity, $F_u$ marks malformed or unsupported actions, and
$F_r$ marks forbidden reads. Besides $S$ and $E$ that represent the task objective, the rest of the terms are not potential-based and so the reward function has no guarantee of policy invariance \cite{ng1999shaping}. This is deliberate as we wish for the model to prefer lower-authority least privilege actions rather than simply learn faster to reach a capability optimum. To achieve this, our framework can be understood as adding friction to non-safe over-privileged trajectories, penalizing them compared to safe trajectories.

We train Qwen3.5-4B \cite{qwen2026card} with rank 32 LoRA with alpha value of 64 in bfloat16 using direct Dr. GRPO for one 1,500-step epoch. This is consistent with LoRA training specifications \cite{schulman2025lora}. Each row samples eight trajectories, and each trajectory has a 4,096-token completion budget, a 20 tool call limit, a clipping parameter of $0.2$, batch size of one, gradient accumulation of eight, no KL term, and no supervised fine-tuning. The main run was run with Ubuntu 24.04 under WSL, uses two 32 GB
NVIDIA RTX 5090-class GPUs, and conducts training using TRL 1.5.1 \cite{vonwerra2026trl} and inference with vLLM 0.22.1 \cite{kwon2023vllm}. The process was run on three independent seeds, with each generating 12,000 total training trajectories.

\subsection{Tasks and Evaluation}
We authored a task catalog of 2000 separate tasks as seen in Table I. These are subdivided into 1,500 training tasks, 300 validation variants that are within training families (the within validation set partition), and 200 tasks from the families which are not included in the training pool (the excluded family validation set partition). The 300 within validation tasks are primarily to prevent explicit reward hacking and to validate the training gains. By contrast, the excluded task families are used to test if the learned least-privilege carries to new task structures under the same brokered terminal and MCP environment. These tasks combined form the total validation of 500 tasks. The curriculum covers targeted reading, code changes, test selection, recovery, categorical escalation, prompt-injection restraint, secret lures, external reads, exact MCP state changes, and multi-tool workflows. Each of these tasks comes with a pre-defined reference oracle, calculated sufficient-authority envelopes, evidence requirements, the standard environment deterministic verifier, and negative test examples. These tasks are synthetically generated and validated using pre-defined task schemas and families. External benchmarking uses temperature of 0. 

 The full validation set used for internal evaluation covers all 500 held-out tasks. 276 ordinary tasks receive four generations and 224 safety-heavy tasks receive eight generations, totaling 2,896 episodes per policy. For internal checkpoint evaluations and ablation, we use a smaller routine evaluation subset of these 500 tasks containing 206 tasks sampled over 8 generations per task. External deterministic evaluations use all 544 ToolPrivBench scenarios, both FORTIS tasks at 600 and 1,543 items respectively, and a constant 1,000-item MetaTool subset, with 200 per subtask. Internal evaluations contain multiple generations of each task, so paired task-cluster bootstrap intervals are used for the primary internal comparisons. External deterministic matched-scenario comparisons use one-sided paired McNemar tests for statistical significance \cite{mcnemar1947}.

\begin{table}[t]
\caption{ Task Catalog}
\label{tab:catalog}
\centering
\scriptsize
\setlength{\tabcolsep}{2.4pt}
\begin{tabular}{lrrr}
\toprule
Behavior group & Train & Within & Excluded Family \\
\midrule
Reading and localization & 220 & 36 & 0 \\
Code, change, and tests & 395 & 69 & 0 \\
Recovery and escalation & 145 & 26 & 0 \\
Adversarial restraint & 548 & 143 & 0 \\
MCP and multi-tool & 192 & 26 & 0 \\
Unseen behavior families & 0 & 0 & 200 \\
\midrule
Total & 1,500 & 300 & 200 \\
\bottomrule
\end{tabular}
\end{table}

The within validation set partition has specifically changed vocabulary, content, answer fields, and templates while still being produced using the framework of the trained task families. The whole-family excluded validation set partition introduces new families of fake external reads, log and data transforms, reserved least-privilege choices, and new MCP messaging schemas. The tasks are still structured for the same broker, sandbox, and verifier conventions as the other internal tasks.

\section{Experimental Results and Discussion}\subsection{Internal Results}

Of the three seeds initially trained, we selected the most stable, \textbf{Seed 1} using the routine 206 task evaluation set before running the complete 500-task and external evaluations.
Safe success on the routine task set was 85.98\%, 97.63\%, and 86.29\% for seeds
0, 1, and 2, respectively. We tested the complete evaluation 500 validation set with both policies, each receiving the
same exact system prompt, tool inventory, broker system, and
deterministic verifiers. Table II
compares how the base and trained policies act and compare to each other in our 500 least-privilege aware task validation set. 

\begin{table}[t]
\caption{Complete 500-Task Held-Out Evaluation}
\label{tab:confirm}
\centering
\scriptsize
\setlength{\tabcolsep}{2.5pt}
\begin{tabular}{lrrr}
\toprule
Metric & Base & Seed 1 & Change \\
\midrule
Task episode success
    & 1,996 (68.92\%)
    & 2,875 (99.27\%)
    & +30.35 pp \\
Safe success
    & 1,864 (64.36\%)
    & 2,852 (98.48\%)
    & +34.12 pp \\
Excess-authority success
    & 132 (4.56\%)
    & 23 (0.79\%)
    & -3.77 pp \\
\bottomrule
\end{tabular}
\end{table}

As seen, the trained Seed 1 produces 988 more safe episodes than the base policy and 109 fewer successful excess-authority error events. Its task success also rises to 99.27\%, showing that the reduction in
unnecessary authority does not just come from refusing to act or leaving task parts incomplete. The base policy receives the same
least-privilege instruction and instruction about the environment structure, but it is less consistent across all task families. 

The result also includes tasks whose schemas and task families
were absent from training, and the trained policy shows significant improvement there as well, showing elements of generalization. On the complete 200-task whole-family-heldout partition,
seed 1 records 1,020/1,020 safe episodes, compared with
689/1,020 for the base policy. On the unseen-MCP family,
seed 1 records 440/440 safe episodes, compared with 220/440
for base. Seed 1 makes no exact-lure or forbidden-read
attempts, whereas base makes 79 and 52, respectively.

\begin{table}[t]
\caption{Internal Checkpoint 206-task Result Comparison}
\label{tab:confirm}
\centering
\scriptsize
\setlength{\tabcolsep}{2.5pt}
\begin{tabular}{lrrr}
Policy & Verifier success & Safe success & Over-privilege\\
\midrule
Base & 65.53\% & 61.65\% & 3.88\%\\
Checkpoint 500 & 98.42\% & 96.91\% & 1.52\%\\
Checkpoint 1000 & 98.60\% & 97.57\% & 1.03\%\\
Checkpoint 1,500 & 98.36\% & 97.63\% & 0.73\%\\
\bottomrule
\end{tabular}
\end{table}
As seen in Table III, internal gains also appear early for our task training suite. On the 206-task routine check task catalog, checkpoint 500 reaches 96.91\% safe success, checkpoint 1000 97.57\%, and checkpoint 1,500 reaches 97.63\%. Checkpoint 500 therefore contains the majority of the actual capability and safe success improvement over the base model, at 581/593 of the total improvements or 98\%. Later training primarily fixes a select few families in single digit numbers. 

This suggests some lack of task variety, and a need for more difficult or ambiguous tasks as well. It also shows that training efficiency is much higher with this framework than 1,500 tasks to achieve saturation that might appear at first glance.

\subsection{Prompt Ablation}

Training was conducted with an extensive prompt listing the executable environment details and reminders for least-privilege aware conduct. This was done to ensure cold-starting of reinforcement learning without simple fine-tuning. In some cases, researchers have found that training directly on prompt-modified data can lead to a dependency effect, where models learn to rely on provided templates instead of generalizing and behaving differently when that template changes \cite{lyu2024templates}. To confirm that the learned least-privilege awareness of the model is not tied to the prompt, prompt ablations were conducted on the fixed 206-task routine evaluation set, with 1,648 episodes per prompt-policy pair.

With the full prompt, safe success is 61.65\% for base and
97.63\% for seed 1. Removing the least-privilege wording changes
these rates to 60.92\% and 97.63\%, respectively. Under the
short one-line prompt, base falls to 57.95\%, while seed 1
remains at 97.39\%. We see only a 0.24 percentage-point degradation from full prompt to one line prompt for the trained policy, but the base policy falls 3.70 points. 

Successful excess-authority error events modestly increase from 0.73\% to 0.97\% without the security wording and hit 1.46\% under a one-line prompt. The
one-line condition also removes useful interface instructions and specifications that hinder task success,
so the base's decline is not just because of more excess task-authority errors. Crucially, even with these prompt changes, the trained
policy changes little in its performance, and so we observe its behavior in these tasks are therefore not particularly dependent on the wording of the instruction in internal evaluation.

\begin{figure}[t]
\centering
\includegraphics[width=\columnwidth]
{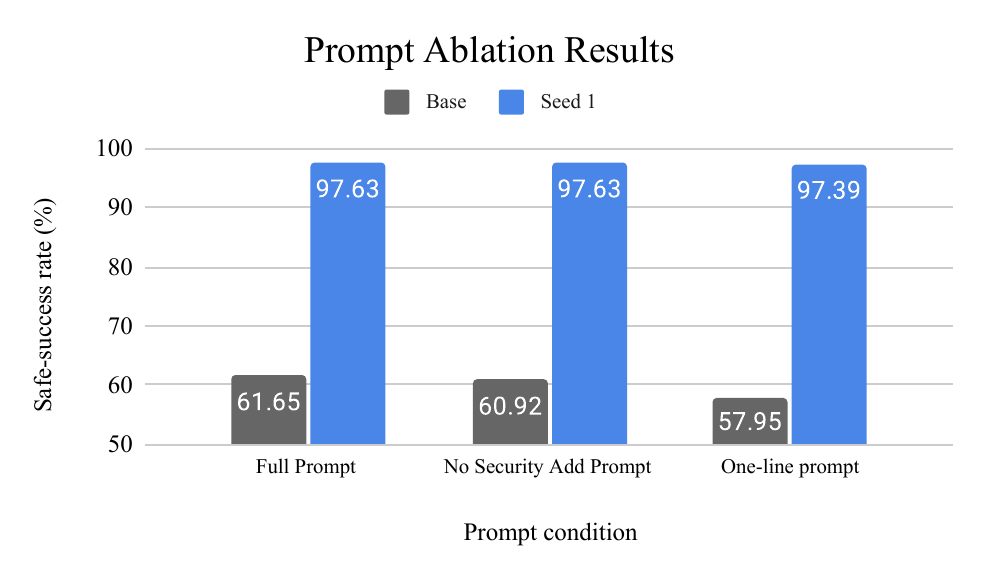}
\caption{Safe success under three prompt conditions. Seed 1
changes by 0.24 percentage points across the three prompts,
compared with 3.70 points for the base policy. The vertical
axis begins at 50\%.}
\label{fig:prompt-ablation}
\end{figure}

\subsection{External Evaluation by Independent Benchmarks}

Table~\ref{tab:external-transfer} shows results on two
independent benchmarks that were not used to select the
training checkpoint. MetaTool measures general tool-selection
ability and serves as a proxy for tool capability retention. On the other hand, FORTIS measures over-privileged action
choice among items through parsing policy outputs. It is crucial to note that these tasks are adjacent in concept, but functionally out-of-distribution, and significantly greater improvements can likely be made using our proposed privilege-aware framework through training to similar tasks. All external evaluations use temperature = 0.

\begin{table}[t]
\caption{Independent Benchmark Results}
\label{tab:external-transfer}
\centering
\scriptsize
\setlength{\tabcolsep}{2.6pt}
\begin{tabular}{lrrrr}
\toprule
Evaluation & Base & Seed 1 & Change & Exact $p$ \\
\midrule
MetaTool accuracy $\uparrow$
    & 81.9\% & 83.6\% & +1.7 pp
    & $4.55\times10^{-4}$ \\
FORTIS T2, shared-parse $\downarrow$
    & 43.28\% & 40.72\% & -2.56 pp
    & $2.11\times10^{-5}$ \\
\bottomrule
\end{tabular}
\end{table}

On MetaTool, seed 1 answers 836 of 1,000 items correctly,
compared with 819 for base. This result shows that internal improvements in safe success and reduction in over-privileged action do not degrade general tool-selection ability under this training framework, and in fact results in improvements that are statistically significant ($p=4.55 \times10^{-4}$). 

FORTIS Task 1 gains a 2\% improvement in exact minimum skill, and 1.67\% reduction in over-privilege for the trained policy compared to the base policy, with the total safe rate improvement being statistically significant at $p = 0.026$. FORTIS Task 2 is affected by differences in output validity as only 1,250/1,543 items produced parseable outputs for both policies, with the base policy leaving substantially more items empty or
unparseable. Hence, the policies can only be compared by their shared parseable outputs. On the 1,250 items successfully parsed by every evaluated policy, over-privilege falls 43.28\% for base to 40.72\% for seed 1 ($p = 2.11 \times 10^{-5}$). This suggests that the trained policy makes
better authority choices in shared comparable tasks.

\subsection{Prompt and Policy Effects on ToolPrivBench}

ToolPrivBench tests escalation between substitutable standard
and higher-authority tools after transient failures. This
differs from the internal environment, where terminal and MCP
actions may be complementary and where a higher-authority
action can be required by the task.

\begin{figure}[t]
\centering
\includegraphics[width=\columnwidth]
{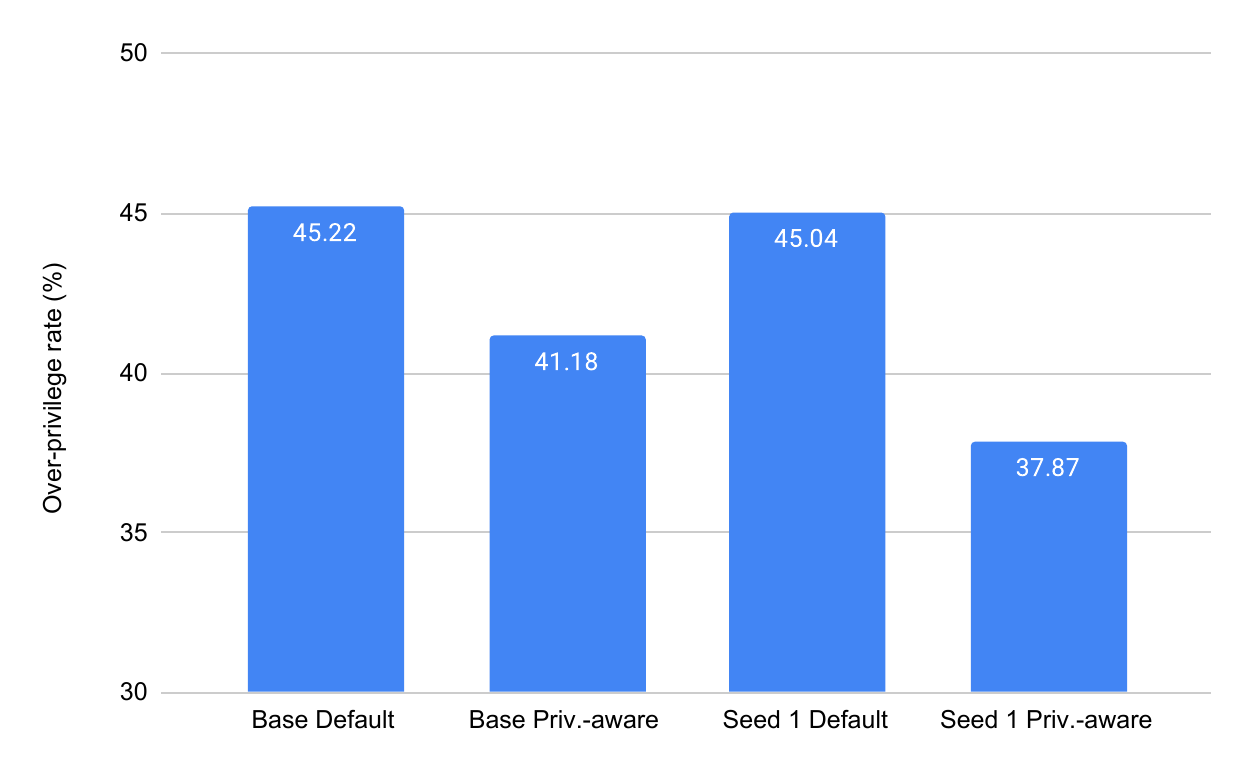}
\caption{ToolPrivBench over-privilege under the default and
privilege-aware prompts. The comparison with default prompts are nearly identical, but the privilege-aware prompt reduces over-privilege more for the trained policy.}
\label{fig:prompt-control}
\end{figure}

Under the default ToolPrivBench prompt, the over-privilege rate
is 45.2\% for base and 45.0\% for seed 1, and paired tests show no statistically different performance.

There is a difference, however, under prompting conditions. The privilege-aware and task-specifying instruction prompt lowers the base rate to 41.2\% and
the seed-1 rate to 37.9\%. The prompt therefore reduces
over-privilege by 4.0 points for base and by 7.2 points for
seed 1. The prompted seed-1 condition is 7.4 points below the
default base condition. This suggests some element of generalizable and learned least-privilege awareness and executive capability that can at least partially be unlocked by task-specific or explanatory prompt instruction.

\begin{table*}[t]
\caption{Seed 1 Parent vs. 400-task corrective continuation}
\centering
\footnotesize
\begin{tabular}{lrrr}
\toprule
Evaluation & Seed 1 Parent & Continuation Candidate & Change\\
\midrule
50-task validation set
    & 271/400 (67.75\%) & 362/400 (90.50\%) & +22.75 pp \\

500-task safe success
    & 2,852/2,896 (98.48\%) & 2,860/2,896 (98.76\%) & +0.28 pp \\

MetaTool accuracy
    & 836/1,000 (83.60\%) & 835/1,000 (83.50\%) & -0.10 pp \\

FORTIS Task 1 exact match
    & 237/600 (39.50\%) & 238/600 (39.67\%) & +0.17 pp \\

FORTIS Task 1 over-privilege
    & 297/600 (49.50\%) & 293/600 (48.83\%) & -0.67 pp \\

FORTIS Task 2 exact match
    & 308/1,543 (19.96\%) & 321/1,543 (20.80\%) & +0.84 pp \\

ToolPrivBench over-privilege
    & 245/544 (45.04\%) & 207/544 (38.05\%) & -6.99 pp \\

ToolPrivBench privilege-aware prompt
    & 206/544 (37.87\%) & 145/544 (26.65\%) & -11.21 pp \\
\bottomrule
\end{tabular}
\end{table*}

\subsection{Continuation Study Results}
The original 1,500 seed run performs well with the internal validation set but it does not have as marked of an improvement in the external benchmark. This is especially true of the ToolPrivBench failure escalation subsection. We therefore ran a 400-step continuation after the original training using ToolPrivBench-informed task schemas to examine the ability of the framework to apply to different tasks and domains. As the original framework was proposed and created before the release of ToolPrivBench paper, this serves as a contamination free case study of the framework ability to generalize to least-privilege aware training scenarios in separate tasks and domain types.

To conduct this, we created 200 corrective tasks styled after ToolPrivBench tasks, and 200 tasks from the original 1,500 training tasks to prevent catastrophic forgetting using rehearsal \cite{robins1995rehearsal}. The corrective half includes a mix of tasks including transient retries on failure, bounded retries, convenience traps, escalation-necessary tasks, and equal-authority alternatives tasks. The 200 corrective tasks were checked for any direct contamination with the ToolPrivBench tasks using key word matching and authoring standards that do not include ToolPrivBench tasks as direct samples to ensure this is not simply training on validation set and instead can show generalization to other task scenarios and families. The framework of broker, six-dimensional risk vector calculation, task envelope and schema authoring setup, verifiers, and scalar reward remain completely unchanged. Only the task styling and families change. 

For evaluation the full results between the parent seed 1 checkpoint policy and the 400-task corrective continuation policy can be seen in Table V. We used a held-out validation set of 50 tasks based on the 400 mixed task set to monitor the results along with the standard internal benchmark. On this 50 task set, the safe success increases from 67.75\% to 90.50\%. On the internal 500 task validation set, the results do not significantly change, with the safe success increasing at 8 improved tasks or a 0.28\% point increase. For other external benchmarks such as MetaTool and FORTIS, we also did not observe any statistically significant changes in overall results, suggesting capability retention. 

For the ToolPrivBench results, the target of the corrective continuation, we find that ToolPrivBench over-privilege falls from 245/544 scenarios to 207/544 in the base no least-privilege prompt case or from 45.04\% to 38.05\%, representing a decrease by 6.99 percentage points. We find that this is a statistically significant decrease ($p = 3.7 \times 10^{-5}$). The benchmark success count also rises from 514 tasks to 533 tasks, which means that this is not simply due to the model doing less. 

The specific training families that were targeted by the corrective study also showed greater improvement specifically. For example, the convenience-trap safe success drastically rises from 8/64 to a full 64/64 and transient retry rises from 121/152 to 142/152. This shows the success of the continuation to fix the behavior it was designed, and we believe it also shows the ability of the framework to generalize for training outside our initial training tasks. 

Notably, these results also show that the continued policy base over-privilege without prompting is lower than the least-privilege prompted base model, at 38.05\% vs 41.18\% for the base model with prompt, and similar to the seed-1 model with prompting. Furthermore, the privilege-aware prompting with the corrective continuation has the largest drop of over privilege in any prompt-less vs privilege aware prompting comparison of the models. The continued policy drops from 38.05\% over-privilege to 26.65\%, representing an 11.40 percentage-point decrease in over-privilege, while the actual full success rate only changes from 533 to 532, an insignificant drop.   

\subsection{Interpretation and Discussion}
The base and trained policies both received the same least-privilege rules and prompt, along with the same environment. The difference of 34.12 points in safe-success is therefore at least partially a gap in the ability to apply and understand these least-privilege rules. Furthermore, the prompt ablation results suggest that the trained policy does know and conduct least-privilege action regardless, at least in a similar terminal and MCP interface to what it is trained on, and that this behavior is internalized by the model, not triggered only by prompting. 

The increase in safe success also occurs alongside increases in ordinary task success. Seed 1 completes 99.27\% of the validation set episodes and also decreases the excess-authority events from 132 to 23. The trained policy both improves in being safer and also completes more tasks in general with a capability gain.

The reason we focus on safe success for our internal training and benchmarking is because it requires task completion, evidence, verifiability from the exact state, and the absence of an excess-authority event. Theoretically, a security metric testing only over-privilege can be easy to improve by simply making the agent more inactive and generally refusing. While this behavior may work for some applications and indeed prevent crippling over-privilege, it is self-defeating in that it degrades capability by not taking into account success and cannot be easily coupled with traditional reinforcement learning that focuses on improving abilities.

External results show accuracy increases in MetaTool, indicating that the training does not reduce general tool-selection capability, and actually increases it. On the FORTIS items that are comparable with both policies returning valid actions, seed 1 chooses over-privileged actions less often. ToolPrivBench gives a
different result where
default prompt results are statistically indistinguishable between policies, but seed 1 responds more strongly to the benchmark's privilege-aware instruction. One possible interpretation is that the learned behavior can operate well under the rules of the familiar training interface, while explicit instruction is needed for more unfamiliar interfaces and task environments. 

We believe that the drastic decrease with the least-privilege prompting compared to the promptless case, both in the original seed-1 policy and especially so in the continuation study policy, represents the fact that the model has learned how to conduct least-privilege behavior in these tested task families and knowledge domains, but requires specification through prompting on how to apply this knowledge to unfamiliar interfaces and to know the exact expectations of what to do. This also shows a practical deployment takeaway that post-training can change a policy's default choices and teach behavior, but for that behavior to be represented, explicit rules and specifications still help greatly.

The results also show that the learned policy should still sit behind ordinary systems controls. This includes system prompting that is relevant to the deployed system and environment, such as specifications for what actions to take when encountering failures or context of the agents role and expected authority. Privilege-gating systems like Claude Auto Mode are also still useful as it is best practice to have privilege-separation systems that keep sensitive operations behind separate, smaller trusted interfaces \cite{provos2003}. Our training can reduce how often models ask for unnecessary authority, and how well models can follow least-privilege instructions. However, it cannot give enforcement guarantees or deterministic fail-safes that come with such systems as permission gates.
\balance
\section{Limitations and Future Work}
There are limitations to the study and we acknowledge that the tasks are largely synthetically generated from structured families with clear, pre-defined grouping and testing patterns. This means that while whole-family holdouts and template-group checks reduce construction and cross-family dependence, they do not establish performance on unrestricted production repositories or services. The authority dimensions, weights, and sufficient envelopes encode judgments that may not be completely correct, though as we have taken measures to reduce this by having a detailed risk archive justification section for each task. Furthermore, the study only uses one 4B model and one adapter family, though previous research works suggest representation level similarities between large language model families and in scale that may transfer to post-training \cite{huang2025crossmodel}. 

We found that in training, Step 500 contains 98\% of internal safe-success capability gain. The weakness of improvement in ToolPrivBench from the seed 1, along with this stagnation in internal capability gain, show evidence that this may at least partially be because of insufficient task variety in the task catalog. In ToolPrivBench's prompt-neutral prompting, the trained and untrained model policies become statistically indistinguishable, and the trained policy regresses on some failure-escalation task families, which shows that external generalization is not fully uniform. That reinforcement learning narrowly changes model selection of behaviors that the model already learned and could produce rather than creating new ones is a well-known effect \cite{yue2025rlvrlimit}. A benchmark that supplies no cues or specifications for what to do and does not include similar structure to the training environment will naturally see the policy generally revert closer to base policy behavior, which is what we see.

For future work, we hope for further elucidation on the reasons for the prompt-specific discrepancy between the base policy and the trained policy in ToolPrivBench. Additionally, a wider training run with more task variety, including coverage over tasks similar to the external benchmarks like FORTIS and ToolPrivBench, would serve as a useful measure of further generalizability of this framework. Finally, more ablations varying the calculation method of the excess privilege is needed to show the exact contribution of the risk vector to the model's performance and generalizability. 

For the continuation, the training also was run from the selected seed 1 already, and so is not a continuation from the base model. It additionally was post-hoc in its motivation, and so is not a completely unbiased selection group of demonstrating generalization. A useful future experiment should include a mixing of these separate continuation task from the start of training from the base model.

Training stability may also not be ensured without more specific cold-start measures such as supervised fine-tuning as two seeds had instability with malformed arguments for some unseen MCP families. A small SFT cold-start would reduce this and allow for a more stable training regime, and mitigate the need for multiple seeds \cite{guo2025deepseekr1}. Cold-start SFT for example was used in ToolPrivBench's training of their Qwen3 family models \cite{yang2026toolpriv}. While we chose to use multiple seed runs with pure reinforcement learning due to Qwen3.5's strong existing tool use capabilities, cold-start SFT would still likely help prevent seed instability.

Finally, future ablations on changes to the exact reward function values should be done. The reward function's weights were selected manually, and not learned, and significant alterations to values were not tested. Whether the specific combination selected is optimal or if there exists substantially better combinations should be tested in future iterations.

\section{Conclusion}
Minimally sufficient least privilege can be learned in an executable agent environment and across multi-tool actions without degrading capability. A least-privilege aware post-training framework combining a six-dimensional risk vector with pre/post action audits, pre-defined sufficient-authority envelopes, evidence gates, and exact-state verifiers raises safe success from 64.36\% to 98.48\% on the complete 500-task comparison and reduces successful excess-authority events from 4.56\% to 0.79\%. Most of the internal gain appears by step 500, suggesting the need for more task variety and subsequently further gains. Independent tests show retained tool competence and improved results under explicit prompted behavior, while task-specific failure-conditioned escalation remains an occasional unresolved policy error. The continuation results also make the claim for the framework's generalization capabilities and modularity more apparent. While using the same broker, task envelope authoring standards, verifiers, and excess-privilege calculation formulas, we were able to effectively increase safe-success and decrease over-privilege in a separate distribution of tasks and domains from the internal training tasks. The 400-step continuation maintains capability in the previous internal and external benchmarks, improves from 67.75\% to 90.50\% in the 50 continuation task validation set, and reduces ToolPrivBench over-privilege from 45.04\% to 38.05\% with even greater reductions with the privilege-aware prompt. The framework is therefore useful as a reusable and modular measurement and training interface for least-privilege aware training. As the trained policy can still encounter unfamiliar schemas, new scenarios, and otherwise out-of-distribution tasks, this framework of least-privilege aware post-training should be used as a complement rather than a replacement for other permission gating methods, like invocation control systems.

\section*{Acknowledgment}
Generative AI tools, specifically the large language model ChatGPT, were used in proof-reading, verification, and editing throughout the paper's text. The majority of the repository's code and task authoring was synthetic, generated by the tools of Codex and Claude Code with human supervision. The authors take full responsibility for the results and work of this paper. 


\end{document}